\PassOptionsToPackage{unicode}{hyperref}
\PassOptionsToPackage{hyphens}{url}
\PassOptionsToPackage{dvipsnames,svgnames,x11names}{xcolor}
\documentclass[
]{article}
\usepackage{amsmath,amssymb}
\usepackage{lmodern}
\usepackage{iftex}
\ifPDFTeX
  \usepackage[T1]{fontenc}
  \usepackage[utf8]{inputenc}
  \usepackage{textcomp} % provide euro and other symbols
\else % if luatex or xetex
  \usepackage{unicode-math}
  \defaultfontfeatures{Scale=MatchLowercase}
  \defaultfontfeatures[\rmfamily]{Ligatures=TeX,Scale=1}
\fi
\IfFileExists{upquote.sty}{\usepackage{upquote}}{}
\IfFileExists{microtype.sty}{% use microtype if available
  \usepackage[]{microtype}
  \UseMicrotypeSet[protrusion]{basicmath} % disable protrusion for tt fonts
}{}
\makeatletter
\@ifundefined{KOMAClassName}{% if non-KOMA class
  \IfFileExists{parskip.sty}{%
    \usepackage{parskip}
  }{% else
    \setlength{\parindent}{0pt}
    \setlength{\parskip}{6pt plus 2pt minus 1pt}}
}{% if KOMA class
  \KOMAoptions{parskip=half}}
\makeatother
\usepackage{xcolor}
\usepackage{color}
\usepackage{fancyvrb}

\DefineVerbatimEnvironment{Highlighting}{Verbatim}{commandchars=\\\{\}}
\newenvironment{Shaded}{}{}

\newcommand{\CommentTok}[1]{\textcolor[rgb]{0.38,0.63,0.69}{\textit{#1}}}

\newcommand{\ControlFlowTok}[1]{\textcolor[rgb]{0.00,0.44,0.13}{\textbf{#1}}}

\newcommand{\DecValTok}[1]{\textcolor[rgb]{0.25,0.63,0.44}{#1}}

\newcommand{\ImportTok}[1]{\textcolor[rgb]{0.00,0.50,0.00}{\textbf{#1}}}

\newcommand{\KeywordTok}[1]{\textcolor[rgb]{0.00,0.44,0.13}{\textbf{#1}}}
\newcommand{\NormalTok}[1]{#1}
\newcommand{\OperatorTok}[1]{\textcolor[rgb]{0.40,0.40,0.40}{#1}}

\usepackage{longtable,booktabs,array}
\usepackage{calc} % for calculating minipage widths
\usepackage{etoolbox}
\makeatletter
\patchcmd\longtable{\par}{\if@noskipsec\mbox{}\fi\par}{}{}
\makeatother
\IfFileExists{footnotehyper.sty}{\usepackage{footnotehyper}}{\usepackage{footnote}}
\makesavenoteenv{longtable}
\providecommand{\tightlist}{%
  \setlength{\itemsep}{0pt}\setlength{\parskip}{0pt}}
\NewDocumentCommand\citeproctext{}{}
\NewDocumentCommand\citeproc{mm}{%
  \begingroup\def\citeproctext{#2}\cite{#1}\endgroup}
\makeatletter
 \let\@cite@ofmt\@firstofone
 \def\@biblabel#1{}
 \def\@cite#1#2{{#1\if@tempswa , #2\fi}}
\makeatother
\newlength{\cslhangindent}
\newlength{\csllabelwidth}
\newenvironment{CSLReferences}[2] % #1 hanging-indent, #2 entry-spacing
 {\begin{list}{}{%
  \setlength{\itemindent}{0pt}
  \setlength{\leftmargin}{0pt}
  \setlength{\parsep}{0pt}
  \ifodd #1
   \setlength{\leftmargin}{\cslhangindent}
   \setlength{\itemindent}{-1\cslhangindent}
  \fi
  \setlength{\itemsep}{#2\baselineskip}}}
 {\end{list}}
\usepackage{calc}

\ifLuaTeX
\usepackage[bidi=basic]{babel}
\else
\usepackage[bidi=default]{babel}
\fi
\babelprovide[main,import]{american}
\def\languageshorthands#1{}
\ifLuaTeX
  \usepackage{selnolig}  % disable illegal ligatures
\fi
\IfFileExists{bookmark.sty}{\usepackage{bookmark}}{\usepackage{hyperref}}
\IfFileExists{xurl.sty}{\usepackage{xurl}}{} % add URL line breaks if available
\hypersetup{
  pdftitle={asdex: Automatic Sparse Differentiation in JAX},
  pdfauthor={Adrian Hill, Guillaume Dalle},
  pdflang={en-US},
  colorlinks=true,
  linkcolor={Maroon},
  filecolor={Maroon},
  citecolor={Blue},
  urlcolor={Blue},
  pdfcreator={LaTeX via pandoc}}

\title{asdex: Automatic Sparse Differentiation in JAX}

\definecolor{c53baa1}{RGB}{83,186,161}
\definecolor{c202826}{RGB}{32,40,38}

\usepackage[affil-it]{authblk}
\usepackage{orcidlink}
\author[1,2%
  \ensuremath\mathparagraph]{Adrian Hill%
    \,\orcidlink{0009-0009-5977-301X}\,%
    }
\author[3%
  ]{Guillaume Dalle%
    \,\orcidlink{0000-0003-4866-1687}\,%
    }

\affil[1]{BIFOLD -- Berlin Institute for the Foundations of Learning and
Data, Berlin, Germany%
  }
\affil[2]{Machine Learning Group, Technical University of Berlin,
Berlin, Germany%
  }
\affil[3]{LVMT, ENPC, Institut Polytechnique de Paris, Univ Gustave
Eiffel, Marne-la-Vallée, France%
  }
\affil[$\mathparagraph$]{Corresponding author: %
  hill@tu-berlin.de %
}
\date{8 October 2026}

\begin{document}
\maketitle

\section{Summary}\label{summary}

Many tasks in scientific computing and machine learning require the
Jacobian or Hessian matrix of a function. Automatic differentiation (AD)
computes these derivatives to machine precision
(\citeproc{ref-baydin2018ad_survey}{Baydin et al., 2018};
\citeproc{ref-blondel2026book}{Blondel \& Roulet, 2026};
\citeproc{ref-griewank2008book}{Griewank \& Walther, 2008}), but
materializing a dense \(m \times n\) Jacobian requires \(n\)
forward-mode or \(m\) reverse-mode AD passes, one per column or row. For
a large class of functions, each output depends on only a few inputs,
making the derivative matrix \emph{sparse}. Automatic sparse
differentiation (ASD) exploits this structure in four steps
(\citeproc{ref-hill2025illustrated}{Hill et al., 2025}):
\emph{detection} of the input-agnostic sparsity pattern, \emph{coloring}
of a graph to group columns or rows that can share an AD pass,
\emph{compressed differentiation} to compute a compressed derivative
matrix with one AD pass per color, and finally \emph{decompression} into
the original sparsity pattern. The number of colors, and hence of AD
passes, is often independent of the problem dimension: a banded Jacobian
with \(b\) contiguous bands, for instance, only ever requires \(b\)
colors, regardless of its size.
\texttt{asdex}\footnote{\url{https://github.com/adrhill/asdex}} offers
the first
standalone ASD toolkit in the popular JAX
(\citeproc{ref-bradbury2018jax}{Bradbury et al., 2018}) ecosystem. With
\texttt{asdex.jacobian} and \texttt{asdex.hessian}, it provides sparse
drop-in replacements for \texttt{jax.jacobian} and \texttt{jax.hessian}.

\section{Statement of need}\label{statement-of-need}

Sparse derivative matrices arise in nonlinear systems of equations,
second-order optimization algorithms, sensitivity analysis, and many
other applications. The target audience for \texttt{asdex} is
researchers and practitioners in scientific machine learning who wish to
leverage JAX's performance, JIT compilation, and accelerator support.

JAX's built-in \texttt{jacfwd}, \texttt{jacrev}, and \texttt{hessian}
functions materialize \emph{dense} derivative matrices, performing one
AD pass per input or output dimension. This standard approach quickly
hits two main limitations: a \emph{memory} bottleneck, if the dense
matrix is too large to store, and a high \emph{computational} cost, as
the number of AD passes scales with the matrix dimension. Computing a
sparse Jacobian (\citeproc{ref-curtis1974sparse_jacobian}{Curtis et al.,
1974}) or Hessian (\citeproc{ref-coleman1984hessian}{Coleman \& Moré,
1984}; \citeproc{ref-powell1979hessian}{Powell \& Toint, 1979}) with ASD
alleviates both of these hurdles by combining more efficient matrix
storage with a faster way to fill it. See Gebremedhin et al.
(\citeproc{ref-gebremedhin2005what_color}{2005}) and Griewank \& Walther
(\citeproc{ref-griewank2008book}{2008, Chapter 8}) for thorough reviews
of this topic.

\section{State of the field}\label{state-of-the-field}

ASD initially matured in low-level programming languages such as Fortran
and C++, but we restrict our review to high-level languages, which
machine learning research favors for the rapid prototyping workflows
they enable.

A general-purpose ASD implementation in a high-level, open-source
language recently appeared in Julia
(\citeproc{ref-bezanson2017julia}{Bezanson et al., 2017}), through the
combination of \texttt{DifferentiationInterface.jl}
(\citeproc{ref-dalle2025di_paper}{Dalle \& Hill, 2026}),
\texttt{SparseConnectivityTracer.jl}
(\citeproc{ref-hill2025sct_paper}{Hill \& Dalle, 2025}) for sparsity
pattern detection, and \texttt{SparseMatrixColorings.jl} (SMC)
(\citeproc{ref-montoison2025revisiting_coloring}{Montoison et al.,
2025}) for coloring. These packages, to which the present authors
contributed, now serve as core sparse differentiation infrastructure for
downstream software such as \texttt{NonlinearSolve.jl}
(\citeproc{ref-pal2026nonlinearsolve}{Pal et al., 2026}). The speedups
they enable are documented in Hill \& Dalle
(\citeproc{ref-hill2025sct_paper}{2025}).

No general-purpose equivalent exists for JAX, although there are a few
prototypes. We list them in the table below with their main features:

\begin{longtable}[]{@{}
  >{\raggedright\arraybackslash}p{(\linewidth - 4\tabcolsep) * \real{0.2467}}
  >{\raggedright\arraybackslash}p{(\linewidth - 4\tabcolsep) * \real{0.2533}}
  >{\raggedright\arraybackslash}p{(\linewidth - 4\tabcolsep) * \real{0.4933}}@{}}
\toprule\noalign{}
\begin{minipage}[b]{\linewidth}\raggedright
Package
\end{minipage} & \begin{minipage}[b]{\linewidth}\raggedright
Detection
\end{minipage} & \begin{minipage}[b]{\linewidth}\raggedright
Coloring
\end{minipage} \\
\midrule\noalign{}
\endhead
\bottomrule\noalign{}
\endlastfoot
\texttt{sparsejac}\footnote{\url{https://github.com/mfschubert/sparsejac}}
& None & Distance-1 (from \texttt{networkx}) \\
\texttt{sparsediffax}\footnote{\url{https://github.com/gdalle/sparsediffax}}
& None & Distance-2 \& star (from SMC) \\
\texttt{jax-nansparse}\footnote{\url{https://github.com/nardi/jax-nansparse}}
& \texttt{NaN} tracing & None \\
\texttt{jax2sympy}\footnote{\url{https://github.com/johnviljoen/jax2sympy}}
& \texttt{jaxpr} symbolic conversion & None \\
\texttt{asdex} & \texttt{jaxpr} tracing & Distance-2 \& star (based on
SMC) \\
\end{longtable}

A \texttt{jaxpr}-based approach to sparsity pattern detection has been
outlined by Simpson (\citeproc{ref-simpson2024sparse_jax_blog}{2024}).
Parts of the ASD pipeline have also been reimplemented internally by
application-specific libraries. The sparse linear solver
\texttt{JAX-AMG}\footnote{\url{https://github.com/jx-wang-s-group/JAX-AMG}}
(\citeproc{ref-liu2026jax_amg}{Liu et al., 2026}) combines
\texttt{jaxpr}-based sparsity detection with a parallel greedy
distance-1 coloring, while the finite element framework
\texttt{tatva}\footnote{\url{https://github.com/smec-ethz/tatva}}
(\citeproc{ref-pundir2026tatva}{Pundir et al., 2026}) pairs
\texttt{jaxpr}-based detection with greedy distance-2 coloring from its
companion package \texttt{tatva-coloring}.

We built \texttt{asdex} rather than contributing to the general-purpose
packages above because none of them combined the ``proper'' sparsity
detection paradigm (\texttt{jaxpr} tracing) with the state-of-the-art
coloring techniques (distance-2 and star colorings). Additionally, many
of the alternatives listed above were made public only recently (we
became aware of the concurrent works \texttt{JAX-AMG} and \texttt{tatva}
as we were writing up this paper).

\section{Software design}\label{software-design}

\subsection{Core features}\label{core-features}

\texttt{asdex} mirrors the main stages of ASD as distinct, composable
components.

\textbf{Sparsity detection.} \texttt{asdex} performs abstract
interpretation on a \texttt{jaxpr}\footnote{\url{https://docs.jax.dev/en/latest/jaxpr.html}}
(short for \emph{JAX expression}), JAX's intermediate representation of
a function. Abstract interpretation propagates index sets through each
\texttt{jaxpr} primitive to determine which inputs influence which
outputs, yielding a conservative pattern that may contain false
positives but never misses a nonzero. Working at the \texttt{jaxpr}
level, the detector reuses JAX's own program representation and
naturally handles functions built from arbitrary JAX primitives. It also
makes second-order detection free: because \texttt{jax.grad(f)} produces
an ordinary JAX function with a \texttt{jaxpr} of its own, and the
Hessian of a scalar-valued function \texttt{f} is the Jacobian of its
gradient, \texttt{asdex} obtains Hessian patterns by applying the same
first-order detector to \texttt{jax.grad(f)}, with no separate
second-order detector to implement or maintain.

\textbf{Coloring.} The detected pattern gives rise to a graph coloring
problem, which \texttt{asdex} solves approximately using standard greedy
algorithms. For Jacobians, a distance-2 coloring of a bipartite
row-column graph partitions columns (forward mode) or rows (reverse
mode) into structurally orthogonal groups (with no overlapping nonzeros)
(\citeproc{ref-gebremedhin2005what_color}{Gebremedhin et al., 2005}). By
default, \texttt{asdex} picks whichever mode requires fewer colors. For
Hessians, a star coloring exploits the symmetry of the matrix to further
reduce the number of colors
(\citeproc{ref-gebremedhin2007acyclic_star}{Gebremedhin et al., 2007}).
In both cases, the resulting groups are encoded in a seed matrix.

\textbf{Compressed differentiation and decompression.} The compressed
matrix stems from parallelized (\texttt{jax.vmap}-ed\footnote{By
  specifying a \texttt{chunk\_size} in \texttt{asdex}, the number of
  products computed in parallel can be limited via \texttt{jax.lax.map},
  bounding memory usage.}) Jacobian-vector or vector-Jacobian products,
evaluated against the seed matrix. All nonzero coefficients are obtained
in this way and then efficiently scattered back into the original sparse
format required for the Jacobian or Hessian.

\subsection{Performance and interface}\label{performance-and-interface}

\textbf{Preparation.} A central design choice separates one-time
\emph{preparation} (detection and coloring, which depend on input shapes
but not values) from repeated \emph{evaluation} (compressed
differentiation and decompression) of the sparse derivative. In an
iterative solver or a training loop, preparation is amortized, and only
the cheap compressed AD pass is repeated, echoing the preparation
mechanism of \texttt{DifferentiationInterface.jl}
(\citeproc{ref-dalle2025di_paper}{Dalle \& Hill, 2026}). While detection
and coloring are not JAX transformations themselves, the prepared
derivative function they produce is an ordinary JAX function, so it
composes with \texttt{jax.jit} and \texttt{jax.vmap} and runs on CPU,
GPU, and TPU backends.

The API deliberately stays close to JAX's own and matches its semantics:
where a dense Jacobian is written as \texttt{jax.jacobian(f)},
\texttt{asdex.jacobian(f,\ x)} differs in only two aspects:

\begin{itemize}
\tightlist
\item
  it takes a sample input \texttt{x} for preparation
\item
  it returns a function computing a sparse matrix (defaulting to JAX's
  \texttt{BCOO} format)
\end{itemize}

\begin{Shaded}
\begin{Highlighting}[]
\ImportTok{import}\NormalTok{ jax}
\ImportTok{import}\NormalTok{ asdex}

\KeywordTok{def}\NormalTok{ f(x):}
    \ControlFlowTok{return}\NormalTok{ (x[}\DecValTok{1}\NormalTok{:] }\OperatorTok{{-}}\NormalTok{ x[:}\OperatorTok{{-}}\DecValTok{1}\NormalTok{]) }\OperatorTok{**} \DecValTok{2}

\NormalTok{x }\OperatorTok{=}\NormalTok{ jax.numpy.zeros(}\DecValTok{1000}\NormalTok{)}
\CommentTok{\# Preparation: detect sparsity \& color, return jit{-}able function}
\NormalTok{jac\_fn }\OperatorTok{=}\NormalTok{ jax.jit(asdex.jacobian(f, x))}
\CommentTok{\# Evaluation: cheap compressed AD passes and decompression}
\NormalTok{J }\OperatorTok{=}\NormalTok{ jac\_fn(x)}
\end{Highlighting}
\end{Shaded}

Beyond this core, \texttt{asdex} mirrors the breadth of JAX's derivative
API. Like JAX's own transformations, it accepts arbitrary
\texttt{PyTree} inputs and outputs rather than only flat vectors.
Multiple arguments (with the \texttt{argnums} keyword, enabling partial
differentiation) and auxiliary outputs (with the \texttt{has\_aux}
keyword) are supported as well. Users can further supply known sparsity
patterns, reuse colorings across sessions, limit their memory usage, and
decompress results into the dense and sparse formats of their choice,
including those from \texttt{jax.experimental.sparse}, \texttt{numpy},
and \texttt{scipy.sparse}. The package is developed openly under the MIT
license. Its documentation spans a tutorial, how-to guides, an API
reference, a contributor's guide, and continuously published benchmarks
tracking detection, coloring, and evaluation performance.

\section{Research impact statement}\label{research-impact-statement}

\texttt{asdex} was inspired by JAX issue \#1032\footnote{\url{https://github.com/jax-ml/jax/issues/1032}},
a feature request for sparse Jacobian and Hessian support, which has
remained open since 2019. This issue centralizes most of the related
discussions and shows that there is sustained interest in such a feature
in the JAX ecosystem. As for \texttt{asdex} itself, it has already been
applied to accelerate the computation of Hessians in machine learning
interatomic potentials (\citeproc{ref-langer2026truncated}{Langer et
al., 2026}) and has been integrated into \texttt{splineax}\footnote{\url{https://github.com/nardi/splineax}},
a sparsity-focused extension of the popular linear and least-squares
library \texttt{lineax} (\citeproc{ref-rader2023lineax}{Rader et al.,
2023}).

\section{AI usage disclosure}\label{ai-usage-disclosure}

Generative AI tools (Anthropic's Claude Opus 4.5 to 5.5 and Fable
5 models) were used in developing the \texttt{asdex} software and
writing this manuscript. All AI-generated code and text were
reviewed and verified by the authors. The design of \texttt{asdex} is
based directly on prior work by the authors that was written without
generative AI tools, namely the peer-reviewed Julia packages
\texttt{DifferentiationInterface.jl}
(\citeproc{ref-dalle2025di_paper}{Dalle \& Hill, 2026}),
\texttt{SparseConnectivityTracer.jl}
(\citeproc{ref-hill2025sct_paper}{Hill \& Dalle, 2025}), and
\texttt{SparseMatrixColorings.jl}
(\citeproc{ref-montoison2025revisiting_coloring}{Montoison et al.,
2025}), from which it inherits its algorithmic approach.

\section{Acknowledgements}\label{acknowledgements}

Adrian Hill gratefully acknowledges funding from the German Federal
Ministry for Research, Technology and Space under the grant BIFOLD26B.
We thank Alexis Montoison for his work on the Julia ASD ecosystem,
Marcel Langer and Denis Korolev for their feedback on \texttt{asdex}, as
well as Jonathan Brodrick, John Viljoen and Johanna Hafner for
productive discussions around JAX.

\clearpage
\section*{References}\label{references}
\addcontentsline{toc}{section}{References}

\protect\phantomsection\label{refs}

\end{document}